\documentclass[prl,aps,twocolumn,showpacs]{revtex4}
\usepackage{epsfig}
\usepackage{amssymb}
\usepackage{amsmath}
\usepackage{graphicx}
\usepackage{amsfonts}
\usepackage{color}

\begin{document}

\author{
T.~E.~Shanygina$^{\bigtriangledown\diamond}$\/\thanks{e-mail: tatiana.shanygina@gmail.com}, Ya.~G.~Ponomarev$^{\diamond}$,  S.~A.~Kuzmichev$^{\diamond}$, M.~G.~Mikheev$^{\diamond}$, S.~N.~Tchesnokov$^{\diamond}$, O.~E.~Omel'yanovskii$^{\bigtriangledown\dag}$, A.~V.~Sadakov$^{\bigtriangledown\dag}$, Yu.~F.~Eltsev$^\bigtriangledown$, A.~S.~Dormidontov$^\bigtriangledown$,
V.~M.~Pudalov$^{\bigtriangledown\bigtriangleup}$, A.~S.~Usol'tsev$^{\bigtriangledown\bigtriangleup}$,  E.~P.~Khlybov$^{\Box\dag}$
}
\address{
$^{\bigtriangledown}$P.~N.~Lebedev Physical Institute RAS, 119991 Moscow, Russia\\
$^{\diamond}$Department of Low Temperature Physics and Superconductivity, Moscow State University, 119991 Moscow, Russia\\
$^\dag$International Laboratory of High Magnetic Fields and Low Temperatures, Wroclaw, 53-421, Poland\\
$^{\bigtriangleup}$Moscow Institute of Physics and Technology, 141700 Moscow, Russia\\
$^{\Box}$Institute for High Pressure Physics RAS, Troitsk, Moscow district, 142190, Russia
}

\title{Observation of Multi-Gap Superconductivity in GdO(F)FeAs by Andreev Spectroscopy}

\begin{abstract}
We have studied current-voltage characteristics
of Andreev contacts in polycrystalline GdO$_{0.88}$F$_{0.12}$FeAs samples with bulk critical temperature $T_c$ = (52.5 $\pm$ 1)\,K using break-junctions technique.
The data obtained cannot be described within the single-gap approach and suggests
the existence of a multi-gap superconductivity in this compound. The large and small superconducting gap values  estimated at $T = 4.2$K are
$\Delta_L$ = $10.5\pm 2$ meV and $\Delta_S$ = $2.3 \pm 0.4$\,meV,
respectively.
\end{abstract}

\pacs{74.70.Xa, 74.20.Fg, 74.25.-q, 74.45.+c, 74.62.Dh}

\date{\today}

\maketitle

Novel superconducting compounds of 1111 family based on rare-earth oxypnictides REOFeAs (RE = La, Sm, Gd etc.) \cite{kamihara06,kamihara08} are currently in the focus of research interest. Some of their features such as layered structure and spatial separation of the carrier reservoir layers and the superconducting pairing layers are similar to those of cuprates. However, many other properties differ substantially  and promise new interesting physics \cite{reviews}. At present, the key issues under investigations are the effect of various types of doping, pairing mechanism, symmetry of the order parameter, quasiparticle energy spectrum, and the superconducting energy gap(s).

The stoichiometric
compounds of the 1111-family are antiferromagnetic metals with spin density wave ground state \cite{AFM}. Partial deficiency of oxygen or fluorine  substitution for  oxygen induces superconductivity in the FeAs-layers. Replacement of  rare-earth elements also affects the superconducting critical temperature, ${T_c}$.
 In particular, ${T_c}$ of Gd-based oxypnictide may be lowered by partial replacement of Y for Gd
 \cite{Y} or gained up by introducing Th instead of Gd \cite{Th}. ${T_c}$=56 K found in Gd$_{0.8}$Th$_{0.2}$OFeAs compound is today  the highest  one for iron-based superconductors.

According to band structure calculations, the total density of states at the Fermi level $N(0)$ is formed mainly by Fe 3d-states \cite{sadovskii_DOS,theory_6,theory_7}. As shown in Ref.~\cite{Kuchinskii}, the ${T_c}$ values for different
iron-based superconductors correlate with $N(0)$, thus giving support to the BCS-like coupling
in these compounds.

The theoretically  calculated  Fermi surface for 1111-system
\cite{theory_1, theory_2, theory_3}
consists of quasi-two-dimensional (2D) hole sheets centered at the $\Gamma$ point and two electron sheets at the $M$ points of the first Brillouin zone. Within the so called minimal two-band model, these four bands may be considered as two effective 2D bands \cite{theory_4, theory_5}.
Correspondingly, many of
the  available theoretical and experimental data indicate
 that iron-based layered materials
  are   multiband superconductors with $s$-type symmetry of the order parameter \cite{reviews}.
  Knight shift measurements in 1111-class compounds  \cite{knight_1111} have proven unambiguously the spin-singlet type pairing in these materials.
  Several data were reported in favor of  $s^\pm$ \cite{zhou08,chen10-flux_jumps} or $s^{++}$  order parameter symmetry \cite{mazin_0901.4790}, making
the experimental situation regarding 1111-compounds uncertain.

The magnitude and structure of the superconducting gap  $\Delta$ is  intimately related to the pairing mechanism.
ARPES measurements are not sensitive enough to resolve unambiguously such fine details as $\Delta$, on the scale of a few meV, that makes this  parameter accessible nearly exclusively from point contact spectroscopy, such as scanning tunneling spectroscopy (STS), tunneling-  and point-contact Andreev reflection (PCAR) spectroscopy (the latter in the regime of SN-, or symmetrical SNS-junctions). The available experimental reports are however rather inconsistent for 1111 class compounds \cite{chen10-Delta, millo08, wang08, pan08}, even  for the most intensively studied SmO(F)FeAs.
Various types of conclusions have been reported including d-wave like, single
gap-like, and multi-gap behavior.

The ambiguity of the experimental information is partly due to an inevitable inhomogeneity of the 1111-type polycrystalline samples  and lack of large size 1111-type single-crystals suitable for these
measurements. Another cause for the divergency of the point contact spectroscopy data  is inherent in those experimental techniques, where  the sample surface is not cleaved in high vacuum or cryogenic environment.
In order to resolve the experimental ambiguity, evidently, novel sets of  comprehensive experimental data are needed, which would comprise self-consistency check, substantial statistics and provide local probing at various points of the in-situ
cleaved surfaces.

Here we report  the superconducting gap  measurements in nearly optimally doped GdO$_{0.88}$F$_{0.12}$FeAs samples by SNS Andreev spectroscopy using the break-junction technique \cite{muller}. Until now these measurements have not been done  for Gd-1111, an analogue to Sm-1111 with approximately the same ${T_c}$$~\sim 53$K.
The break junction technique opens a nice opportunity to
prepare in helium atmosphere, at liquid $^4$He temperatures, clean surfaces forming Andreev contact. Another advantage is a possibility of fine mechanical readjusting the contact during experiment, that enables to
 collect multiple data from different local areas of the same sample.
Using this method we have unambiguously detected  the presence of two superconducting gaps, whose best fit values averaged over about 30 spectra are $\Delta_L=10.5\pm 2$meV and $\Delta_S=2.3\pm 0.4$\,meV at $T=4.2$K.

Polycrystalline samples GdO(F)FeAs were prepared by high pressure synthesis \cite{khlybov_JETPL}.
The  chips of high purity Fe, and powders of single-phase FeF$_3$,  Fe$_2$O$_3$, and  GdAs  were mixed together
in the nominal ratio
and  pressed into pellets of 3mm diameter and 3mm height.
The pellets were
placed in boron nitride crucible and synthesized  at pressure of 50\,kb  and temperature 1350$^\circ$C during 60\,min.
The X-ray diffraction pattern averaged over the sample area  showed
a polycrystalline compound
with a dominating desired 1111-phase (with lattice parameters $a= 3.902(2)${\AA}, $c= 8.414(5)${\AA} )
and an admixture of incidental FeAs and Gd$_2$O$_3$
phases.
The subsequent local EDS analysis (JSM-7001FA) has revealed that the  incidental phases are concentrated in grains of about 1mkm size which are scattered in the bulk majority phase. This fact opens a possibility  to  probe properties of the true majority phase using local techniques, such as   PCAR.
Superconducting properties of the samples were tested by measurements of temperature dependence of ac-magnetic susceptibility and resistivity $R(T)$. Both  showed a sharp superconducting transition in our  polycrystalline samples with $T_c\approx 52.5$K (the latter value was defined at a maximum of $dR(T)/dT$-curve). Figure~1 shows typical temperature dependence of resistance and its derivative.

\begin{figure}[hc]
\begin{center}
\includegraphics[width=.5\textwidth]{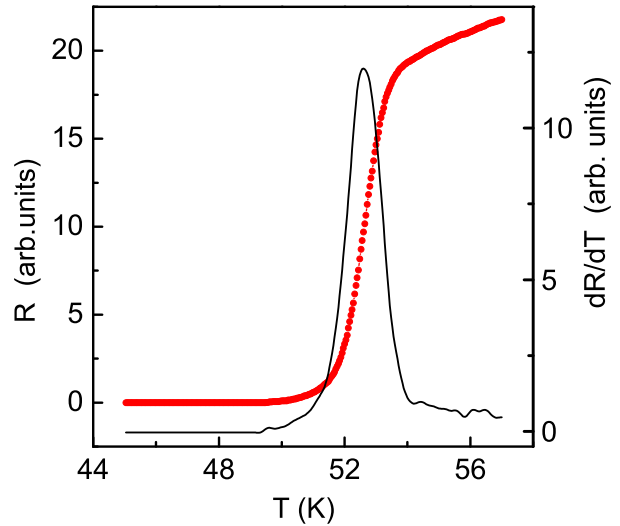}
\caption{Fig.~1. Superconducting transition for polycrystalline GdO$_{0.88}$F$_{0.12}$FeAs sample measured prior a microcrack formation
 (dots). The bulk
$T_c$ = (52.5 $\pm1$)\,K was determined at a maximum in $dR(T)/dT$-curve (solid line).}
\end{center}
\label{fig:F1}
\end{figure}

For point-contact spectroscopy we used
two methods: (i) multiple Andreev reflections spectroscopy of individual superconductor-constriction-superconductor Sharvin-type contacts \cite{andreev, kummel, sharvin} and (ii) intrinsic Andreev specrtoscopy of stack contacts that usually exist due to the presence of steps and terraces on clean cryogenic cleaves in layered crystals.

Thin plates of about $2\times1\times0.12$ mm$^3$ in size were cut from the synthesized pellets. At room temperature, the plate-like sample was
mounted onto an elastic bronze holder and the two current and two potential leads were attached to the sample  by liquid In-Ga alloy. The holder with the sample was placed in the measuring cell   and cooled down to 4.2\,K. A microcrack in the sample was generated by precise bending the sample holder at $T = 4.2$K using a micrometric screw.

Current-voltage dependence, $I(V)$, and its derivative, $dI(V)/dV$,
were measured  automatically using the 16-bit AT-MIO-16X (National Instruments) digital board.
The amplitude of a low-level 820 Hz modulation voltage at potential leads of a sample was maintained
stable  using a lock-in nanovoltmeter (operated as null-
detector) and a computer controlled digital bridge with a proportional-integral-derivative feedback signal.
As a result, the differential conductance of a contact was proportional to the amplitude of the ac feedback current through the contact.

Figure~2 represents $I(V)$, $dI(V)/dV$ and $d^2I(V)/dV^2$
characteristics for individual Andreev (SNS) break-junction in polycrystalline GdO$_{0.88}$F$_{0.12}$FeAs sample measured at $T = 4.2$K.
The observed experimental $IV$-curves are typical for the clean classical SNS-contacts with excess-current characteristics \cite{kummel, aminov}, therefore, the theoretical model of K\"{u}mmel et al. \cite{kummel} is supposed to be applicable to our break-junctions.
According to the K\"{u}mmel model, the $IV$-characteristics at low bias voltages should show a subharmonic gap structure (SGS) with a series of dips in the dynamic conductance $dI(V)/dV$ at bias voltages
\begin{equation}
V_n = 2\Delta /en,
\end{equation}
with an integer $n = 1, 2...$,  due to multiple AR effect. For a two-gap superconductor, two independent SGSs corresponding to the large $\Delta_L$ and small $\Delta_S$  gaps are anticipated.

\begin{figure}[hc]
\begin{center}
\includegraphics[width=.45\textwidth]{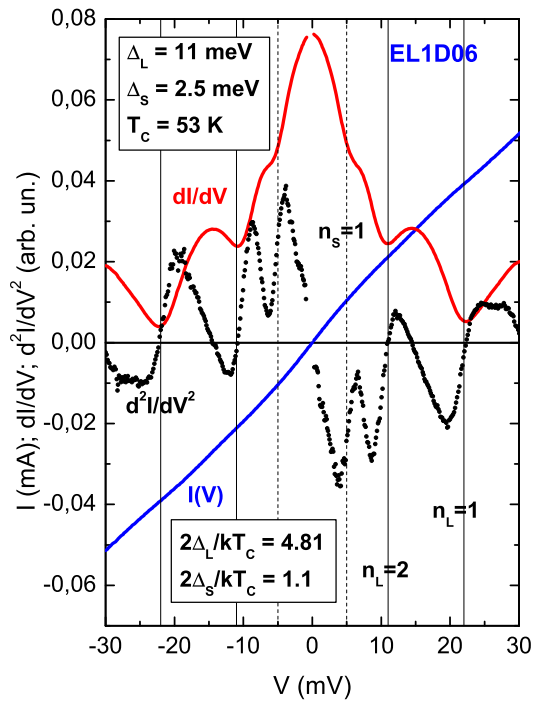}
\caption{Fig.~2.
$I(V)$, $dI(V)/dV$
and $d^2I(V)/dV^2$-curves  for a single SNS-contact 1D06
at $T = 4.2$K. Background (a  polynomial function) is subtracted. The  set of dips in the differential conductance  at bias voltages $V_{nL} = 2 \Delta_{L}/en$ (vertical solid lines)
determines the energy of the large
superconducting gap, $\Delta_L\approx 11$\,meV. Peculiarities  on the $dI(V)/dV$ and $d^2I(V)/dV^2$-curves,  marked by dashed lines, indicate the presence of small superconducting gap, $\Delta_S \approx 2.5$\, meV.}
\end{center}
\label{fig:F2}
\end{figure}

The dips labeled on Figure~2 as $n_L =  1$ and 2 on SGS reflect  the large gap; they are marked with vertical solid lines.
The singularities
shown by vertical dashed lines cannot be attributed to the large gap and, therefore, may reflect the
existence of a small gap $\Delta_S \approx 2.5$\, meV.
 Comparing the result for the  large gap (Fig.~2) with Eq.~(1) one can easily obtain  $\Delta_L \approx 11 $\,meV.

By readjusting the contact, we could observe
clear sets of dips on $dI(V)/dV$ curve due to either large-  or small-gaps, or even to both (as shown in Fig.~2). Figure~3  expands  the
small bias range,  where the subharmonics  of the small gap are clearly seen in $dI/dV$ curves for a single SNS-contact EL3D01. For clarity, the smoothly varying  background is subtracted.
Using  expression (1), from the  set of differential conductance dips
on the $dI(V)/dV$ curve we
 determined the energy of
 the small superconducting gap  $\Delta_S = 2.15$\,meV. In Fig.~3, one can also see  two extra features at  $\approx 3$\,mV and 3.7\,mV, which can not be attributed to either large or small gap subharmonics. They may  be caused by excitation of collective  modes and  require additional studies.

\begin{figure}[hc]
\begin{center}
\includegraphics[width=.47\textwidth]{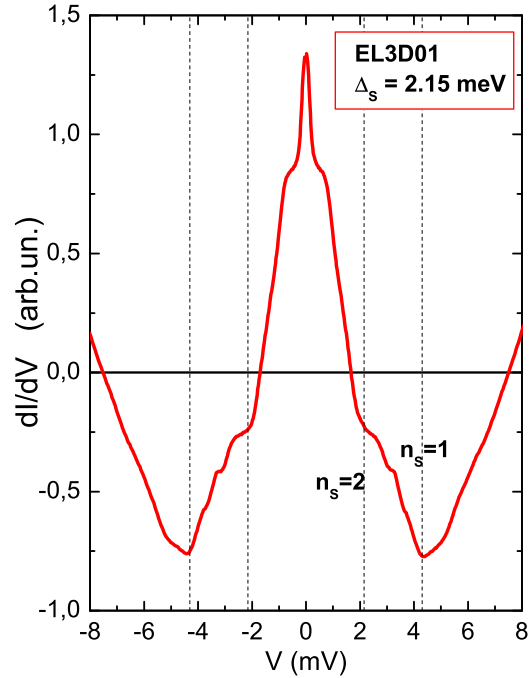}
\caption{Fig.~3. Differential conductance of the SNS contact 3D01 in
GdO$_{0.88}$F$_{0.12}$FeAs sample at $T = 4.2$K. Background is subtracted. Two  differential conductance dips define the energy of the small  superconducting gap
 $\Delta_S
 \approx 2.15$\,meV. The anticipated bias voltages $V_{nS}= 2 \Delta_S/en$ are depicted by vertical dashed lines.}
\end{center}
\label{fig:F3}
\end{figure}

The peculiarities in the $dI(V)/dV$ and $d^2I(V)/dV^2$-characteristics shown in Figs. 2 and 3 clearly manifest the existence of two gaps in our GdO(F)FeAs samples. This behavior is similar to that  observed  earlier  in the multi-band superconductor Mg$_{1-x}$Al$_x$B$_2$   \cite{MgB} and in LaO$_{0.9}$F$_{0.1}$FeAs \cite{ponomarev}
(an analog to our Gd-1111 sample with somewhat lower $T_c\approx 28$K).
The sharpest SGS (like those in Figure~3) may be usually
observed only on $dI(V)/dV$-characteristics of  Andreev contacts of the high quality and of small size,
comparable to the quasiparticles  mean free path (ballistic limit) \cite{sharvin}. For such a case, a
number of the observable gap peculiarities
(up to 4 in some samples)
facilitates
interpretation of the multigap subharmonic structure.

\begin{figure}[hc]
\begin{center}
\includegraphics[width=.45\textwidth]{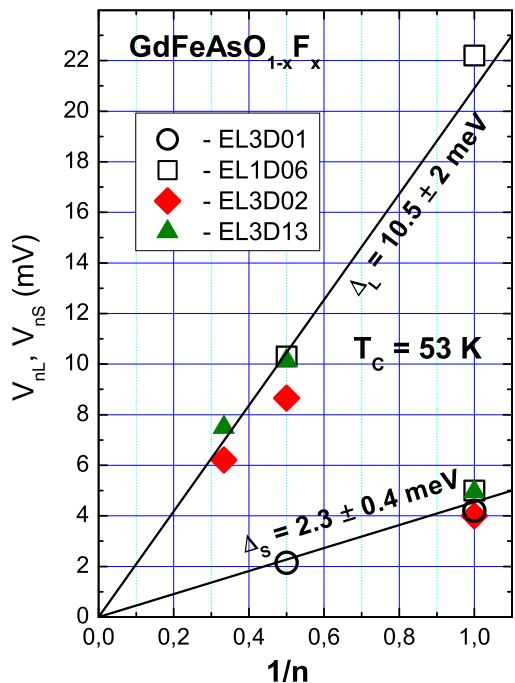}
\caption{Fig.~4. Normalized bias voltages $V_{n} = 2 \Delta_{L,S}/en$ versus $1/n_{L,S}$ for the studied SNS-arrays. The averaged values of the superconducting gaps are  $\Delta_L = (10.5 \pm 2)$meV and $\Delta_S = (2.3 \pm 0.4)$meV. Solid lines are guides to the eye.}
\end{center}
\label{fig:F4}
\end{figure}

Our experimental data are summarized in Figure~4 where the normalized to a single junction bias voltages $V_{nL,S}$ for four microcontacts are plotted versus $1/n_{L,S}$. According to expression (1), such dependences have to
fall onto straight lines passing through zero. This is indeed fulfilled for $\Delta_L$ in all samples;  for $\Delta_S$, this however, could  be verified only in sample EL3D01, because a rich  picture of reproducible features detected at low bias voltages impeded the analysis for other samples.

Based on the  data obtained we
conclude on the existence of   two distinct superconducting gaps with energies  $\Delta_L = (10.5 \pm 2)$\,meV and $\Delta_S = (2.3 \pm 0.4)$meV at $T = 4.2$K in GdO$_{1-x}$F$_{x}$FeAs sample. The reproducibility of two SGSs detected at $dI(V)/dV$-characteristics of various Andreev arrays, formed by the break-junction technique support this conclusion. In some cases we observed  extra features in $dI/dV$-curves signalling the existence of the 3rd,  smaller gap, $\Delta_{SS}\sim 1$\,meV.

Using the determined gap energies and  bulk $T_c = (52.5 \pm 1)$K,
 one can estimate
$2\Delta/k_BT_c$  ratio. For the large gap, our experimental data lead to $2\Delta_L/k_BT_c = (4.8 \pm 1.0)$ that exceeds the standard   BCS value, 3.52, for single-gap superconductors in the weak coupling limit. This fact together with rather
conventional exponent value for Fe isotope effect \cite{isotope} resembles the BCS -model behavior with strong electron-phonon coupling. At the same time, the $2\Delta/k_BT_c$ ratio for the small gap $2\Delta_S/k_BT_c\approx 1.1 \ll 3.52$ suggests that the ``weak'' superconductivity
may be induced by interband coupling, due to $k$-space internal proximity effect between two condensates, where the large gap condensate plays the  ``driving'' role. In particular, similar situation is believed to be realized in MgB$_2$ \cite{MgB} and LaO$_{0.9}$F$_{0.1}$FeAs \cite{ponomarev}.
\begin{table}[h]
\caption{Table 1. Summary of the $\Delta$ values (in meV) measured for 1111-family REO(F)FeAs compounds by point contact Andreev reflection (PCAR), break-junction Andreev reflection (BJ), scanning tunneling spectroscopy technique (STS), and ARPES}

\begin{tabular}{|c||c|c|c|c|c|}
\hline
RE & T$_c$(K) & method & $2\Delta_L$& $2\Delta_S$ & ref.\\
\hline
Gd& 53   & BJ      & 10.5$\pm$2     & 2.3$\pm$0.4  & this work\\
Sm & 53   & ARPES   & 15$\pm$1.5   & no           & \cite{kondo}\\
Sm & 52   & PCAR    & 18$\pm$3     &6.15$\pm$0.45 & \cite{daghero2}\\
Sm & 52   & PCAR    & 19           & 5.7          &\cite{gonelli}\\
Sm & 52   & STS     & 8-8.5        & no           &\cite{millo08}\\
Sm & 51.5 & PCAR    & 20           & 6.6          & \cite{gonelli}\\
Sm & 51   & PCAR    & 10           & 4            & \cite{wang08}\\
Nd & 51   & PCAR    & 14$\pm$1     & 6 $\pm$1     & \cite{miyakawa}\\
Nd & 51   & PCAR    & 12.5$\pm$0.5 & 6.3$\pm$0.3  & \cite{tanaka}\\
Nd & 48   & BJ,STS  & 7-10         & no           & \cite{ekino}\\
Tb & 45   & PCAR    & 8.8          & 5            &\cite{yates}\\
Nd & 45   & PCAR    & 11$\pm$2     & 5 $\pm$1     & \cite{samuely}\\
Sm & 42   & PCAR    & 15$\pm$1     &4.9$\pm$0.5   & \cite{daghero2}\\
Sm & 42   & PCAR    & 6.7$\pm$0.1  & no           & \cite{chen10-flux_jumps}\\
\hline
\end{tabular}
\label{Delta}
\end{table}

The presence of the large superconducting gap characterized by 2$\Delta_L/k_B T_c> 3.52$ in the 1111-family compounds REOFeAs (RE = La, Sm, Nd) was confirmed by tunneling spectroscopy using break-junction technique \cite{ekino}, point-contact Andreev reflection spectroscopy \cite{ chen10-Delta, wang08,  daghero2,  gonelli, miyakawa, tanaka, yates, samuely}, scanning tunneling spectroscopy \cite{millo08, ekino}, and angle-resolved photoemission spectroscopy (ARPES) \cite{kondo} (see Table 1). To the best of our knowledge, there is no other available data  for Gd-1111. Therefore, we
compare in Table 1 our data for Gd-1111 with other data available for Sm-, Nd- and Tb-1111 superconductors with similar $T_c$.
We emphasize rather good agreement between $2\Delta_L/k_B T_c$ values
determined from  our study and those from STS \cite{millo08}, break-junction measurements  \cite{ekino}, and some PCAR-measurements \cite{wang08, tanaka, yates, samuely}.

As to the small gap, there is evidently a sizeable spread in its value ($2 - 5)$meV observed in different experiments.
We can not also exclude that  the spread may be caused by the existence of three-gap superconductivity in 1111-system  with multiple-sheet
 Fermi surface  \cite {theory_3, lebegue}; for the smallest gap, we estimate  $2\Delta_S/k_BT_c\approx 0.5 $.

In conclusion, we have studied the $I(V)$- and $dI(V)/dV$-characteristics at $T = 4.2$K for various  SNS Andreev break-junctions in polycrystalline GdO$_{0.88}$F$_{0.12}$FeAs samples with bulk critical temperatures $T_c \approx 52.5$K. The obtained characteristics do not follow the standard single-gap model behavior. Two clearly observed independent subharmonic gap structures point
at the existence of
two distinct superconducting gaps, $\Delta_L = (10.5 \pm 2)$meV and $\Delta_S = (2.3 \pm 0.4)$meV determined at $T=4.2$K. The estimated
$2\Delta_L/k_BT_c$  ratio exceeds the standard BCS value, 3.52, for single-gap superconductors and weak-coupling limit
while  for the small gap the $2\Delta/T_c$-ratio
  $2\Delta_S/k_BT_c \ll 3.52$.

The authors are grateful to E.~G.~Maksimov,  E.~V.~Antipov, and S.~M.~Kazakov  for valuable discussions and S.~M.~Kazakov  for  help in X-ray diffraction characterization of the samples. The work was partially supported by the Russian Foundation for Basic Research
and by the Russian Ministry for Education and Science.

\end{document}